\documentclass[%
twocolumn,
 amsmath,amssymb,
 prd,
 aps,
longbibliography,
preprintnumbers,
]{revtex4-1}
\pdfoutput=1
\usepackage{graphicx}
\usepackage{dcolumn}
\usepackage{bm}
\usepackage{hyperref}

\usepackage{xcolor, soul}
\definecolor{IITred}{rgb}{0.5,0.05,0.05}

\newcommand{\mev}{\hbox{ MeV}}
\newcommand{\mevc}{\hbox{ MeV}\!/\!c}
\newcommand{\gev}{\hbox{ GeV}}
\newcommand{\gevc}{\hbox{ GeV}\!/\!c}
\newcommand{\tev}{\hbox{ TeV}}
\newcommand{\s}{\hbox{ s}}
\newcommand{\m}{\hbox{ m}}
\newcommand{\eq}[1]{Eqn.\thinspace(\ref{#1})}
\newcommand{\orcid}[1]{\thanks{\href{https://orcid.org/#1}{ORCID: #1}}}
\begin{document}


\title{Notes on Lepton Gyromagnetic Ratios}

\author{Chris Quigg }\email{quigg@fnal.gov}\orcid{0000-0002-2728-2445}\preprint{FERMILAB-FN-1129-T}
\affiliation{Theoretical Physics Department, Fermi National Accelerator Laboratory\\ P.O. Box 500, Batavia, Illinois 60510 USA}

\date{\today}

\begin{abstract}
\centerline{A compendium for outsiders.}
\end{abstract}

\maketitle


\section{\label{sec:origins}Origins}

Dirac's quantum theory of the electron tells us what an electron is~\cite{[][{. See also }]Dirac:1928hu,*9780198520115}:
a particle that carries half a quantum $(\hbar/2)$ of spin angular momentum,  $-1$ unit of electric charge, and magnetic moment $\mu_e$ of minus one Bohr magneton,
$-\mu_{\mathrm{B}} \equiv -\hbar e/2m_e$. Here $\hbar$ is the quantum of action, $-e$ the electron charge, and $m_e$ the electron mass.
The Bohr magneton is defined as the magnitude of the magnetic dipole moment of a point electron orbiting an atom with
one unit ($\hbar$) of orbital angular momentum. [The numerical value is $\mu_\mathrm{B} = 5.788\,381\,8060(17)\times10^{-11}\hbox{ MeV T}^{-1}$~\cite{[][{. This is the source for  parameters not otherwise attributed in this note.}]Zyla:2020zbs}.]
On the classical level, an orbiting point particle with electric charge $e$ and mass $m$ exhibits a magnetic dipole moment given by
\begin{equation}
\vec{\mu}_{L}=\frac{e\hbar}{2 m} \vec{L} .\label{eq:Lmom}
\end{equation}
Dirac's prediction for the electron magnetic moment is thus twice
the value that would arise for a half unit of orbital angular momentum, if that were possible. The ratio $g_e \equiv \mu_e/(-\frac{1}{2}\mu_{\mathrm{B}})$  is called
the gyromagnetic ratio of the electron; the Dirac equation  predicts $g_e = 2$.

 These properties are precisely what is required to account for what Dirac calls
 ``duplexity phenomena,'' the 
observed number of quantum states for an electron in an atom being twice
the number given by the quantum theory of a spinless point particle. A spin-$\frac{1}{2}$ electron with a magnetic moment of one Bohr magneton matches
 precisely what  Uhlenbeck and Goudsmit~\cite{[{}] [{. Their first publication on electron spin is }]UHLENBECK1926, *[][{. Spin is identified as the mysterious fourth quantum number posited by }]Uhlenbeck:1925wb,*1925ZPhy...31..765P} inferred from their study of atomic spectra.

Dirac writes, \emph{The question remains as to why Nature should have chosen this particular
model for the electron instead of being satisfied with the point-charge. One
would like to find some incompleteness in the previous methods of applying
quantum mechanics to the point-charge electron such that, when removed,
the whole of the duplexity phenomena follow without arbitrary assumptions.
In the present paper it is shown that this is the case, the incompleteness of
the previous theories lying in their disagreement with relativity, or, alternatetively,
with the general transformation theory of quantum mechanics. It
appears that the simplest Hamiltonian for a point-charge electron satisfying
the requirements of both relativity and the general transformation theory
leads to an explanation of all duplexity phenomena without further assumption}~\cite{[{What
he understands by ``transformation theory'' is explained in }][{. We might simply say, ``quantum mechanics.''}]doi:10.1098/rspa.1927.0012}.

\section{ Magnetic Moment as A Diagnostic \label{sec:diagnostic}}
\subsection{Generalities\label{subsec:gen}}
Within quantum electrodynamics (QED), a renormalizable local relativistic quantum field theory of photons and electrons, the gyromagnetic ratio of a lepton, $g_\ell$, emerges {unambiguously} from the perturbation expansion. 
An extensive and interesting literature explores what makes $g_\ell = 2$  the ``natural value'' for a structureless point particle of spin-$\frac{1}{2}$~\cite{[{For a systematic presentation, see }][{. Other interesting considerations appear in }]doi:10.1119/1.2345655,*PfisterKing,*FPT}.
 The requirement of good high-energy behavior implies that $g_\ell - 2$ must vanish at tree level for any well-behaved theory~\cite{WeinbergBrandeis}.

Quantum corrections induce a deviation from the Dirac moment that is traditionally expressed as the   magnetic moment anomaly,
 \begin{equation}
 a_{\ell} \equiv \frac{g_{\ell}-2}{2}. \label{eq:anomaly}
 \end{equation}
 The predicted value of $a_\ell$ can be confronted by experiment. The model dependence of the magnetic anomaly makes it an incisive test of QED and a sensitive probe for new-physics contributions~\cite{[{For a helpful introduction to the subject, see }][]Giudice:2012ms}. The interplay between theory and experiment played a decisive role in the development and validation of quantum electrodynamics~\cite{[][{. See in particular Chapter 5, ``The Lamb Shift and the Magnetic Moment of the Electron.''}]9780691033273}.
 
Beyond its role in challenging QED and probing for the virtual influence of undiscovered particles and forces, the anomalous magnetic moment provides stringent constraints on lepton substructure~{\cite{Brodsky:1980zm}}. For that purpose, we may examine the \emph{magnetic anomaly defect,} defined as the difference between theoretical prediction and experimental determination,
\begin{equation}
\delta a_\ell  \equiv a_\ell^{\mathrm{th}} - a_\ell^{\mathrm{exp}} . \label{eq:composa}
\end{equation}
A very simple working hypothesis is that no cancellations or suppression factors due to symmetries in the underlying dynamics account for the  small mass $m_{\ell}$ of the composite lepton itself. Alternatively, so-called chiral models provide more modest constraints. In either case, it is informative to relate the compositeness scale $M^*_\ell$ and the radius $R_\ell$ of the lepton to the magnetic anomaly defect,
\begin{subequations}
\label{compos}
\begin{eqnarray}
\left|\delta a_\ell\right| & = & \frac{m_{\ell}}{M^*_\ell} = m_{\ell} R_{\ell} \quad \hbox{no suppression, or} \label{eq:composb} \\
\left|\delta a_\ell\right| & = & \left(\frac{m_{\ell}}{M^*_\ell}\right)^{\!2} = m_{\ell}^2 R_{\ell}^2 \quad \hbox{chiral model.} \label{eq:composc}
\end{eqnarray}
\end{subequations}
%
 Searches for quark and lepton compositeness in high-energy collisions, reviewed in \S92 of Ref.~\cite{Zyla:2020zbs}, reach above $10\tev$, again assuming no dynamical conspiracies.

\subsection{The Electron}
The electron was the focus of Dirac's theory and the test case for the developing theory of quantum electrodynamics in the late nineteen-forties. It is stable on the time scale of any conceivable experiment; the current bound on the electron lifetime, $\tau_e > 6.6 \times 10^{28}\hbox{ yr}$ at 90\% C.L.~\cite{[][{. Borexino is an exquisitely radiopure liquid scintillation detector  located deep underground at the Gran Sasso Laboratory. The lifetime bound derives from a search for electron decay into a neutrino and a single monoenergetic photon.}]Agostini:2015oze}, greatly exceeds the age of the universe. The electron mass is $m_e = (0.510\,998\,946\,1 \pm 0.000\,000\,003\,1)\mev$. \textsf{CPT} invariance requires that the gyromagnetic ratios of electron and positron be identical: $g_{e^-} = g_{e^+}$.

The anomalous magnetic moment of the electron was discovered in 1947 by Kusch and Foley~\cite{PhysRev.72.1256.2,*PhysRev.74.250}, who inferred
\begin{equation}
a_{e^-} ^{[\mathrm{KF}]}= 0.001\,19 (5)
\end{equation}
from their study of hyperfine structure in gallium atoms. Julian Schwinger showed that the one-loop (${O}(e^2)$) quantum correction to the electron's magnetic moment contributes~\cite{[][{. Schwinger's grave marker in the Mount Auburn Cemetery in Cambridge, Massachusetts, bears the inscription, $\frac{\alpha}{2\pi}$.}]Schwinger:1948iu}\
\begin{equation}
a_e^{[2]} = \frac{\alpha}{2\pi} \approx 0.001\,162, \label{eq:JSae}
\end{equation}
where the numerical value reflects the value of the fine-structure constant as then known, $\alpha = 1/137$.

The two-loop contribution to the gyromagnetic ratio of the electron is also known analytically~\cite{[{The first calculation to include a fermion loop was given by }][{. An important correction leading to the result displayed here is due to }]Karplus:1950zzb,*[][{ and to }]Petermann:1957hs,*Sommerfield:1957zz,*SOMMERFIELD195826}. It is
\begin{subequations}
\label{sommerfield}
\begin{eqnarray}
a_e^{[4]} & = & \frac{\alpha^{2}}{\pi^{2}}\left(\frac{197}{144}+\frac{\pi^{2}}{12}+\frac{3}{4} \zeta(3)-\frac{ \pi^{2}}{2} \ln 2\right)\label{eq:CSaea}\\
 & = & -0.328\, \left(\frac{\alpha}{\pi}\right)^{\!2} \approx -0.000\,001\,77. \label{eq:CSAeb}
\end{eqnarray}
\end{subequations}
Here $\zeta(3) = \sum_{i=1}^\infty i^{-3} = 1.202\,056\,903\,\ldots$ is the Riemann zeta function of 3.
 The sum of \eq{eq:JSae} and \eq{sommerfield} yields the theoretical prediction ca.\thinspace1957 (for a theory of photons and electrons only),
\begin{equation}
a_e \approx a_e^{[2]} + a_e^{[4]} =  0.001\,159\,6.
\end{equation}
%
The contributions of heavier fermion $(f\bar{f})$ bubbles are suppressed by mass ratios. The leading contribution when the mass ratio $(m_e/m_f)$ is small is~\cite{Jegerlehner:2009ry}
\begin{equation}
a_e^{[4:f]} \approx  \frac{1}{45}\,\frac{\alpha^{2}}{\pi^{2}}\left(\frac{m_e}{m_f}\right)^{\!2}, \label{eq:hvy}
\end{equation}
so that $a_e^{[4:\mu]} \approx 0.000\,000\,52\,(\alpha^2/\pi^2)$. A form useful for evaluating fermion bubbles for all values of the mass ratio is given in Ref.~\cite{Passera:2004bj}.

Through heroic work over many decades, the calculation of the electron's magnetic anomaly has been carried out through five loops in QED. The three-loop ${O}(\alpha^3)$ contribution is given in closed form in~\cite{Laporta:1996mq}; evaluated numerically, it gives
\begin{equation}
a_e^{[6]} = (1.181\,241\,456 \ldots)\left(\frac{\alpha}{\pi}\right)^{\!3} .
\end{equation}
 Analytical calculations, reinforced by numerical evaluations, exist through four loops, i.e., up to ${O}(\alpha / \pi)^{4}$, or eighth order in the electron charge $e$.  The coefficient of $(\alpha / \pi)^{5}$ is known from numerical integrations. In addition, the contributions of weak and hadronic interactions have been estimated. 
The predicted value as of 2019 is~\cite{atoms7010028}
\begin{equation}
a_{e}^{\mathrm{th[Cs]}} = 115\,965\,218\,160.6(11)(12)(229) \times 10^{-14}, \label{eq:aoyamaraw}
\end{equation}
where the first two uncertainties are from the tenth-order QED term and the hadronic term, respectively. The third and largest uncertainty comes from the value of the fine structure constant obtained from atom-interferometry measurements of the Cs atom,
$\alpha^{-1}(\mathrm{Cs})=137.035\,999\,046(27)$~\cite{Parker_2018}. A more recent determination using matter-wave interferometry to measure the recoil velocity of a rubidium atom that absorbs a photon, leads to  $\alpha^{-1}(\mathrm{Rb20}) = 137.035\,999\,206(11)$~\cite{Morel2020,*[][{. For an assessment, see }]saida,*Mueller2020}. These two highly precise values  differ by approximately $5.4$ standard deviations~{\cite{[{Ref.~\cite{atoms7010028} quoted an earlier measurement of $\alpha^{-1}(\mathrm{Rb})$: }][] Bouchendira:2010es}}.

To compare theory and experiment, it is efficient to combine the theoretical uncertainties of \eq{eq:aoyamaraw}, thus:
\begin{equation}
a_e^{\mathrm{th[Cs]}} = 115\,965\,218\,161(023) \times 10^{-14} .  \label{eq:aoyamacomb}
\end{equation}
This represents a prediction at the level of 0.23  parts per trillion (ppt) for $g_e$. Adopting instead the 2020 value of $\alpha^{-1}(\mathrm{Rb})$~\cite{Morel2020}, the standard-model prediction is
\begin{equation}
a_{{e}}^{\mathrm{th[Rb20]}}=115\,965\,218\,025.2(95) \times 10^{-14}, \label{eq:LKB2020}
\end{equation}
which carries an uncertainty of 0.1 ppt for $g_e$ and lies $5.5\sigma$ below $a_e^{\mathrm{[Cs]}}$. A smaller value of $\alpha^{-1}$, which is to say a larger value of $\alpha$, implies a larger calculated value of $a_e$. The difference is very closely given by the Schwinger contribution,
\begin{equation}
\frac{\alpha(\mathrm{Cs}) - \alpha({\mathrm{Rb20}})}{2\pi} \approx 0.136(25) \times 10^{-11}, \label{eq:schwdiff}
\end{equation}
a  shift to which modern measurements of $g_e$ are sensitive.

An independent numerical evaluation of the $O(e^{10})$ contribution of diagrams without 
lepton loops has been carried out by Volkov~\cite{[][{. See \S I for the 
consequences for $a_e$ and $\alpha^{-1}$.}]Volkov:2019phy}, with a 
result that differs slightly from the result of Ref.~\cite{atoms7010028}. 
Although the contending values differ by $4.8\sigma$, the implications for 
$a_e$ are not significant  for  current comparisons of theory and 
experiment; a resolution will be needed in the near future.

%
%
Today, experimental determinations of $g_e$ have attained sub-ppt precision---a stunning achievement. 
The evolution of experimental technique up to 1972 is reviewed in~\cite{Rich:1972ru}, which also contains a thorough historical summary of theoretical developments. That chronology recounts the landmark experiments at the University of Michigan that directly observed 
the spin precession of polarized electrons in a region of
static magnetic field. 
This technique culminated in a 3.5 parts-per-billion (ppb) determination of $g_e$~\cite{PhysRevA.4.1341},
\begin{equation}
a_{e^-}^{\mathrm{[Mich]}}=1\,159\,657.7\,(3.5) \times 10^{-9}. \label{eq:umich} 
\end{equation}

Measurements on trapped single electrons led, over time, to a 4.3-ppt determination of $g_e$ at the University of Washington (UW)~\cite{[][{. See also }]PhysRevLett.59.26,*Dehmelt:1990zz,*Brown:1985rh}. The result for the anomalous moment is
\begin{equation}
a_{e^-}^{\mathrm{[UW]}}=1\,159\,652\,188.4\,(4.3) \times 10^{-12}
\end{equation}
The same experiment yields a positron anomaly of
\begin{equation}
a_{e^{+}}^{\mathrm{[UW]}} =1\,159\,652\,187.9\,(4.3) \times 10^{-12},
\end{equation}
validating the prediction of \textsf{CPT} symmetry to a remarkable degree:
\begin{equation}
g_{e^-} / g_{e^+}=1+(0.5 \pm 2.1) \times 10^{-12}. \label{eq:CPTtest}
\end{equation}
These stood as the definitive measurements for nearly two decades.


The most precise published result, from the Harvard University team, is obtained by resolving the quantum cyclotron and spin levels of a single electron suspended for months in a cylindrical Penning trap. This work reaches an uncertainty of 0.28 ppt for $g_{e^-}$~\cite{[][{. The value of $\alpha^{-1}$ inferred from the measurement, based on eighth-order theory, is superseded by the tenth-order result quoted in \eq{eq:alphaae}.}]Hanneke:2008tm}.
Written in the same form as the UW results, it is
\begin{equation}
a_{e^-}^{\mathrm{[H08]}} = 1\,159\,652\,180.73(0.28) \times 10^{-12}. \label{eq:hogae} 
\end{equation}



Comparing the prediction \eq{eq:aoyamaraw} and measurement \eq{eq:hogae}, we find that the magnetic anomaly defect of the electron is
\begin{equation}
\delta a_{e^-}^{\mathrm{[Cs]}} \equiv a_e^{\mathrm{th[Cs]}} - a_{e^-}^{\mathrm{[H08]}} = (88 \pm 37) \times 10^{-14},\label{eq:deltaae}
\end{equation}
with the measurement $2.4\sigma$ below the prediction.  The difference between \eq{eq:LKB2020} and \eq{eq:hogae} yields
\begin{equation}
\delta a_{e^-}^{\mathrm{[Rb20]}}\equiv  a_e^{\mathrm{th[Rb20]}} - a_{e^-}^{\mathrm{[H08]}} = (-48 \pm 30) \times 10^{-14}, \label{eq:deltaaeLKB}
\end{equation}
with the measurement $1.6\sigma$ above the prediction. 
Taking $\left|\delta a_e\right| \lesssim 10^{-12}$ as a rough measure of a potential offset between theory and experiment, we infer from \eq{eq:composb} a compositeness scale $M_e^* \gtrsim 5 \times 10^5\tev$, or equivalently a composite-electron radius $R_e \lesssim 4 \times 10^{-25}\m$. The chiral-invariant \emph{Ansatz}  gives, through \eq{eq:composc}, the more modest limits $M_e^* \gtrsim 500\gev$ and $R_e \lesssim 4 \times 10^{-19}\m$. 

By equating their formal (five-loop) expression for the electron anomaly to \eq{eq:hogae}, Aoyama et al.~\cite{atoms7010028} take $a_{e^-}^{\mathrm{[H08]}}$ as a measure of the fine structure constant,
 \begin{equation}
 \alpha^{-1}\left(a_{e^-}\right)=137.035\,999\,149\,6(13)(14)(330). \label{eq:alphaae}
 \end{equation}
  The uncertainties are from the tenth-order QED term, hadronic term, and  $a_{e^-}^{\mathrm{[H08]}}$, respectively. The inferred value, $\alpha^{-1}\left(a_{e}\right)=137.035\,999\,150(33)$, lies $0.104(43)\times 10^{-6}$ ($2.4\sigma$) above $\alpha^{-1}(\mathrm{Cs})$~\cite{Parker_2018} and $0.056(35) \times 10^{-6}$ ($1.6\sigma$) below $\alpha^{-1}(\mathrm{Rb})$~\cite{Morel2020}, as we anticipate from  Eqns.\thinspace(\ref{eq:deltaae}, \ref{eq:deltaaeLKB}).

Planned improvements in technique by the Gabrielse Research Group at Northwestern University (formerly Harvard)~{\cite{[][{ For general information, see \href{https://cfp.physics.northwestern.edu/gabrielse-group/gabrielse-home.html}{the Gabrielse Lab web site.} See also }]Gabrielse:2019cgf,*PhysRevA.103.022824}} aim at an order-of-magnitude improvement over the precision of $a_{e^-}^{\mathrm{[H08]}}$  and a 150-fold improvement in the accuracy of $g_{e^-}/g_{e^+}$ compared with \eq{eq:CPTtest}.

Should furher studies establish a discrepancy between the standard-model
prediction and the experimental determination of $g_e$, it is of interest
to ask whether a new light $X$ boson, feebly coupled to the electron, might
be responsible.  The NA64 experiment at CERN, which directs a 100-GeV
electron beam onto stationary nuclei has set new exclusion
limits on a sub-GeV boson that decays predominantly into invisible
final states~\cite{[][{.  This article contains an extensive list of useful
references.}]Andreev:2021xpu}.


\subsection{The Muon}
The muon, the second-generation charged lepton, has a mass of $m_\mu =105.658\,374\,5\,(24)\mev$ and a mean life $\tau_\mu = 2.196\,981\,1\,(22) \times 10^{-6}\s$.
The parity-violating decay $\mu^+ \to e^+ \nu_e \bar{\nu}_\mu$ (and charge-conjugate) correlates the muon spin direction and the direction of the emitted positron (or electron),
as first demonstrated by Garwin, Lederman, and Weinrich~\cite{[][{; see also }]Garwin:1957hc,*Friedman:1957mz} at Columbia University's Nevis Cyclotron. In leading approximation, the correlation is given by $(1 -\frac{1}{3}\cos\theta)$, where $\theta$ is the angle between the muon spin and the emitted electron momentum~\cite{[{For a treatment of the muon spin asymmetry through $O(\alpha^2)$ in QED, see }][]Caola:2014daa}.
Garwin, et al.\ were able to exploit this fact in their discovery experiment to constrain the muon's gyromagnetic ratio to lie within 5\% of the $O(\alpha)$ prediction, $2\left(1 + \alpha/2\pi\right) \approx 2.001\,162$ (cf.\ \eq{eq:JSae}). This was quickly followed by an improved measurement in Liverpool, which reported $g_\mu = 2.004 \pm 0.014$ and is notable for an early example of the ``wiggle plot'' that is a feature of later storage-ring experiments~\cite{[][{. The (anomalous) precession frequency of a muon moving through a magnetic field is imprinted on the time distribution of high-energy positrons emitted along the $\mu^+$ spin direction in the form of oscillations on the exponential decay curve characteristic of the (boosted) muon lifetime.}]Cassels_1957}. Subsequent measurements by Garwin and collaborators led by 1960 to the determination $a_\mu  = 0.001\,13^{+0.000\,16}_{-0.000\,12}$, in good agreement with the contemporaneous prediction, $a_\mu^{\mathrm{th:1960}} = 0.001\,16$~\cite{Coffin:1958zz,*Garwin:1960zz}.


Over nearly two decades, three elegant storage-ring experiments at CERN greatly advanced both technique and precision. They reported
\begin{subequations}
\label{CERNsr}
\begin{eqnarray}
a_\mu^{1965} & = & 116\,200\,0\,(5000) \times 10^{-9} \quad~\textrm{\cite{Charpak:1965zz}},  \\
a_\mu^{1972} & = & 116\,616\,0\,(310) \times 10^{-9} \quad~~\,\textrm{\cite{Bailey:1972eu}}, \\
a_\mu^{1979} & = & 116\,592\,4\,(8.5) \times 10^{-9}\quad~~~\textrm{\cite{Bailey:1978mn}}.
\end{eqnarray}
\end{subequations}
The final entry averages the measured anomalous magnetic moments of positive and negative muons, $a_{\mu^{+}}=$ $116\,591\,1(11) \times 10^{-9}$ and $a_{\mu^{-}}=116\,593\,7(12) \times 10^{-9}$.

The progression of \eq{CERNsr} represents a nearly 600-fold improvement through the series of experiments, and the beginning of meaningful constraints on nonstandard contributions to the anomalous moment. The CERN experiments provided a baseline for  later efforts at Brookhaven and Fermilab. 

The muon anomalous magnetic moment reported by Brookhaven experiment E821 value~\cite{[][{. Also see \href{https://www.g-2.bnl.gov}{the E821 web site}. The central value is adjusted to the latest value of $\mu_\mu/\mu_p$ by the Muon $g - 2$ Theory Initiative.}]Bennett:2006fi}
\begin{equation}
a_\mu^{\mathrm{BNL}} = 116\,592\,089\,(63)\times 10^{-11}, \label{eq:BNLamu}
\end{equation}
 set a new standard for precision: its uncertainty of 0.54 ppm  represented a 14-fold improvement compared to the classic CERN measurements. The result was highly consequential, because it suggested a mismatch between theory and experiment. In the words of the Abstract,
 \begin{quote}
 While the QED processes account for most of the anomaly, the largest theoretical uncertainty, $\approx 0.55$ ppm, is associated with first-order hadronic vacuum polarization. Present standard model evaluations, based on 
$e^+ e^-$  hadronic cross sections, lie 2.2--2.7 standard deviations below the experimental result.
\end{quote}
The determination $(g_{\mu^+} - g_{\mu^-})/\langle{g_\mu}\rangle = (-0.11 \pm 0.12)\times 10^{-8}$ is consistent with \textsf{CPT} invariance~\cite{PhysRevLett.92.161802}.

The E821 report stimulated not only a vigorous commerce in candidate interpretations of the putative break in the standard model, but also intensive effort to make the theoretical prediction more secure, largely by scrutinizing the expectations for the hadronic vacuum polarization and the contribution---confirmed to be small---of  light-by-light scattering. Planning for an improved experimental test led to the development of Fermilab Muon $g - 2$ program using the relocated BNL muon storage ring~\cite{Carey:2009zzb}.


A detailed review of theory and experiment through 2009 appears in Ref.~\cite{Jegerlehner:2009ry}.
At two-loop order, the contribution of a muon loop in the photon vacuum polarization may be taken from \eq{sommerfield}.  The electron loop contributes~{\cite{Suura:1957zz,*[][{; and Sommerfield, Ref.~\cite{Karplus:1950zzb}. For a complete evaluation of the electron loop, retaining $m_e/m_\mu$ terms, see }]Petermann:1957ir,*ELEND1966682}
\begin{subequations}
\label{elend}
\begin{eqnarray}
a_\mu^{[4:e]} & = & \frac{\alpha^2}{\pi^2}\left[-\frac{25}{36} + \frac{1}{3}\ln{\frac{m_\mu}{m_e}}+ O\!\left(\frac{m_e}{m_\mu}\right)\right] \\
 & \approx & 1.094\,  \frac{\alpha^2}{\pi^2} .
 \end{eqnarray}
\end{subequations}
Combining with \eq{eq:JSae} and \eq{eq:CSAeb}, we find \begin{equation}
a_\mu = \frac{\alpha}{2\pi} +{0.766}\,\frac{\alpha^2}{\pi^2} + \cdots \approx (0.000\,116\,6). \label{amu4est}
\end{equation}
The QED contributions to the anomaly are now known through five loops ($O(e^{10})$)~\cite{Aoyama:2012wk}.

In support of the experimental program at Fermilab (E989), an international Muon $g-2$ Theory Initiative is working to refine the standard-model prediction. Their consensus, reported in a 2020 White Paper (WP)~\cite{Aoyama:2020ynm}, is
\begin{equation}
a_{\mu}^{\mathrm{[WP:2020]}}=116\,591\,810(43) \times 10^{-11}. \label{eq:WP}
\end{equation}
It differs from the BNL E821 result  \eq{eq:BNLamu} by  $279 (76)$ in the three trailing  places, so lies $3.7\sigma$ below the experimental result.
[The uncertainty in  $a_\mu^{[2]}$ due to the contending measurements of the fine structure constant (cf.\ \eq{eq:schwdiff})  is negligible on the scale of current experimental capabilities.] 

The consensus relies on data-driven dispersion-relation evaluations of the hadronic vacuum polarization~\cite{Davier:2019can,Keshavarzi:2019abf}. Much current activity is devoted to \emph{a priori} evaluations using lattice techniques. 
The path toward a purely theoretical calculation  was laid out by  Blum~\cite{PhysRevLett.91.052001} nearly two decades ago. An independent evaluation is of course desirable; in addition, the lattice calculation has advantages for the separation of QED effects from hadronic corrections and the treatment of isospin breaking. Perhaps half a dozen groups are pursuing this program, aiming to determine the hadronic vacuum polarization contribution with sub-percent precision. Their progress is reviewed in Refs.~\cite{MEYER201946,Aoyama:2020ynm}. 


Fermilab E989 recently reported its first results~\cite{[][{. More information is available at \href{https://muon-g-2.fnal.gov}{the E939 web site.}}]PhysRevLett.126.141801}: 
\begin{equation}
a_{\mu}^{\operatorname{FNAL}}=116\,592\,040\,(54) \times 10^{-11}. \label{eq:e989}
\end{equation}
This 0.46-{ppm} determination of $a_\mu$, within $0.6\sigma$ of the BNL value, is $3.3\sigma$ above the theory consensus. 
Combining the BNL and FNAL measurements gives a 2021 world average,
\begin{equation}
\langle a_{\mu}^{2021}\rangle=116\,592\,061(41) \times 10^{-11}\quad(0.35~\mathrm{ppm}), \label{eq:amu21}
\end{equation}
$4.2\sigma$ above the theory consensus. The support for the BNL central value and the modest strengthening of the offset between theory and experiment is  suggestive that new physics of some sort may be at play. Although measurements of $a_\mu$ are less precise than those of $a_e$, the muon's heavier mass confers greater sensitivity to new-physics effects~\cite{[{It is an interesting challenge to exploit the greater precision of $a_e$ to test interpretations of the $a_\mu$ deviation. See }][]PhysRevA.89.052118}.

It is noteworthy that the observed offset,
\begin{equation}
\langle a_{\mu}^{2021}\rangle - a_{\mu}^{\mathrm{[WP:2020]}} = 251\,(59) \times 10^{-11} \label{eq:offset}
\end{equation}
is not small. It is comparable in size to the electroweak contribution to the anomaly~\cite{Czarnecki:2002nt,*[][{. The corresponding contribution for the electron is $a_e^{\mathrm{[EW]}} = 3.053\,(23) \times 10^{-14}$: }]Gnendiger:2013pva,*Jegerlehner:2017zsb},
\begin{equation}
a_\mu^{\mathrm{[EW]}} = 153.6\,(1.0) \times 10^{-11}. \label{eq:amuew}
\end{equation}

A good story needs a wrinkle, and Muon $g-2$ is no exception. The Budapest--Marseille--Wuppertal Collaboration has presented a new lattice-QCD evaluation of the hadronic vacuum polarization, which enables them to predict~\cite{[][{. See also }]BMWNature,*fodor}
\begin{equation}
a_\mu^{\mathrm{[BMW]}} = 116\,591\,954\,(55) \times 10^{-11}. \label{eq:BMWamu}
\end{equation}
The difference from the consensus value \eq{eq:WP} is subtle but telling. The BMW value
 lies $144\,(70) \times 10^{-11}$ ($2.1\sigma$) above the consensus value and just $107\,(69) \times 10^{-11}$ ($1.6\sigma$) below the experimental world average. If it accurately reflects the standard-model prediction, there is no particular evidence for a discrepancy with experiment. 
 
 Needless to say, theoretical and experimental work continues, with a particular focus on the hadronic vacuum polarization. The BABAR Collaboration~\cite{Druzhinin:2019koy}, the CMD-3~\cite{CMD3} and SND~\cite{Achasov:2020iys} experiments at the Budker Institute in Novosibirsk, and the  Belle-II experiment~\cite{Kou:2018nap} recently commissioned at the KEK Super-B factory all aim at refining the $e^+e^- \to \hbox{hadrons}$ data sets. The MUonE experiment foreseen at CERN aims to gather information in the spacelike region by making precise measurements of  $\mu e$ elastic scattering~\cite{Calame:2015fva,*[][{ For a recent update, see }]Abbiendi:2016xup,*Abbiendi:2020sxw}.

The (experimental average) magnetic anomaly defect with respect to the consensus prediction~\cite{Aoyama:2020ynm} is
\begin{equation}
\delta a_\mu = 251(59) \times 10^{-11} \label{eq:delamu}
\end{equation}
which reflects the $4.2\sigma$ mismatch between calculation and world average measurement. If we take
$\left|\delta a_\mu\right| \lesssim 3 \times 10^{-9}$, we estimate $M_\mu^* \gtrsim 3.5 \times 10^4\tev$ and $R_\mu \lesssim 5.6 \times 10^{-24}\m$ using \eq{eq:composb} or $M_\mu^* \gtrsim 1.9\tev$ and $R_\mu \lesssim 10^{-19}\m$ using \eq{eq:composc}.

The analysis reported in Ref.~\cite{PhysRevLett.126.141801} is based on 6\% of the planned  E989 data sample. The current uncertainties are 434 ppb statistical and 157 ppb systematic. By the summer of 2022, the collaboration expects to report on their Run 2 and Run 3 data sets, increasing to four times the current sample (approximately 10 times the BNL E821 sample) and reducing the experimental error by a factor of two. At that point, they foresee a systematic uncertainty at the 100 ppb level. The ultimate goal is to record and analyze 20 times the BNL sample, leading to a further reduction of a factor of two in uncertainty.

At the Japan Proton Accelerator Research Complex in Tokai, J-PARC experiment E34~\cite{[][{. More information is available at \href{https://g-2.kek.jp/portal/}{the experiment's web page}.}]10.1093/ptep/ptz030} will employ a very different technique, using a $300\mevc$
 reaccelerated thermal muon beam with 50\% polarization. The beam will be  injected vertically into a  solenoid storage ring with 1 ppm local magnetic field uniformity for the muon storage region with an orbit diameter of 66 cm. Compare the Brookhaven and Fermilab experiments, with muon momentum of $3.09\gevc$ and orbit diameter $14.224\m$.
The precision goal for $a_{\mu^+}$  is a statistical uncertainty of 450 ppb, similar to the statistical weight of the BNL and FNAL-2021 samples, and a systematic uncertainty less than 70 ppb.

\subsection{The Tau Lepton}
The third charged lepton, $\tau$, has a mass 
 $m_\tau = 1\,776.86\,(12)\mev$, so its anomalous magnetic moment should have a heightened sensitivity to quantum corrections from heavy particles.  
 In particular, $a_\tau$ is more sensitive than $a_\mu$ to heavy ``new physics'' by a ratio of $(m_\tau/m_\mu)^2 \approx 280$.
By the same logic, $a_\tau$ should also be more sensitive to strong-interaction contributions.
 
 In common with the muon, the tau lepton has parity-violation decays, here
\begin{subequations}
\label{taubrs}
\begin{eqnarray}
\tau^- & \to & \mu^- \bar{\nu}_\mu \nu_\tau  \quad (17.39  \pm 0.04 ){ \%}    \\
\tau^- & \to & e^- \bar{\nu}_e \nu_\tau \;\quad (17.82 \pm 0.04 ) \%
\end{eqnarray}
\end{subequations}
that analyze its spin direction. However, experiments similar to those carried out for the muon are made challenging by the short lifetime of the $\tau$ lepton, $\tau_\tau=(290.3 \pm 0.5) \times 10^{-15}\s$.
For that reason, conceiving a technique to measure $a_\tau$ demands original thinking~\cite{[{See Problem 1.6 of }][{; second edition, (Princeton University Press, Princeton, 2013).}]Quigg:1983gw}.\\

A recent theoretical prediction, within the standard-model paradigm, is~\cite{Keshavarzi:2019abf,[{For an earlier estimate, see }][]Eidelman:2007sb}
\begin{equation}
a_{\tau}^{\mathrm{th}}= 117\,717.1\,(3.9) \times 10^{-8} , 
\end{equation}
slightly greater than $a_e$ and $a_\mu$. What information we have from experiment is indirect, derived from limits on anomalous $\sigma_{\mu\nu}$ couplings of $\tau$ to electromagnetism. The \emph{Review of Particle Physics}~\cite{Zyla:2020zbs} takes as the best current constraint a limit from the DELPHI experiment at LEP~\cite{[][{. Compare the value ${a}_{\tau}=0.004 \pm 0.027 \pm 0.023$ from the rate of hard photons in $e^+e^- \to \tau^+\tau^-\gamma$ reported by }] Abdallah:2003xd,*[][{. See also }]Acciarri:1998iv,*[][{ for the result $-0.068<a_{\tau}<0.065$.}]Ackerstaff:1998mt} that  derives from the measured cross section for the reaction  $e^+e^- \to e^+e^-\tau^+\tau^-$, which is to say $\gamma\gamma \to \tau^+\tau^-$: 
\begin{equation}
-0.052 < a_\tau < 0.013 \hbox{ at 95\% CL}, 
\end{equation}
or $a_\tau = -0.018 \pm 0.017$. While this result is consistent with the Dirac value, $g_\tau =2$, it does not meaningfully test either the standard-model prediction or the presence of unexpected quantum corrections.

How could we do better? A reasonable near-term goal might be to reach the level of the Schwinger contribution, \eq{eq:JSae}. A spin-precession experiment seems out of the question, so improved measurements of the rates for two-photon production of tau pairs, or of hard-photon emission accompanying tau-pair production, merit consideration. An example of original thinking is a proposal to study $\gamma\gamma \to \tau^+\tau^-$ in ultraperipheral heavy-ion collisions at the Large Hadron Collider, aiming at a three-fold improvement on the DELPHI constraint~\cite{[][{. Exploratory---but competitive---measurements are reported in \href{https://cds.cern.ch/record/2803742?ln=en}{CMS Physics Analysis Summary HIN-21-09,} ``Observation of $\tau$ lepton pair production in ultraperipheral nucleus--nucleus collisions,'' and in ATLAS Collaboration, }]Beresford:2019gww,*ATLAS:2022ryk}. The BELLE-II experiment~\cite{Kou:2018nap} aims to record 45 billion tau pairs, which will enable new searches for a (\textsf{P}- and \textsf{T}-violating) electric dipole moment and constraints on $a_\tau$~\cite{Fael:2013ij,*Eidelman:2016aih}. What are the prospects for future electron--positron colliders: the International Linear Collider~\cite{ILCsite}, FCC-ee~\cite{Abada:2019lih}, CEPC~\cite{CEPCStudyGroup:2018ghi}, CLIC~\cite{CLICsite}, etc.?

%
\vspace*{30pt}

\section{\label{sec:sansPAMD}{\lowercase{$g=2$} without  Dirac?}}
%
%

Dirac's statement of purpose and the elegant coherence of his results suggest
that the gyromagnetic ratio $g=2$ for the electron is a consequence of special relativity~\cite{[{This conviction is reinforced by Dirac's debt to }][]Thomas:1926dy,*[][{. For a reminiscence, see }]Thomas:1927yu,*Thomas:1982cj}. And yet, 
following an argument given by Feynman~\cite{9780805325010}, 
modern quantum mechanics textbooks show that the canonical result can be recovered by simple manipulations
within the framework of  nonrelativistic quantum mechanics, without explicitly invoking relativity~\cite{[][{, pp.~78--79. }]sakurai1967,*[][{, where Dirac's derivation is recast in modern vector notation. See also the survey by }]Shankar1994,*doi:10.1119/1.17066}.


Using the vector identity
\begin{equation}
\boldsymbol{\sigma} \cdot \mathbf{X} \medspace\boldsymbol{\sigma} \cdot \mathbf{Y}=\mathbf{X} \cdot \mathbf{Y}+i \boldsymbol{\sigma} \cdot \mathbf{X} \times \mathbf{Y},
\end{equation}
where $\boldsymbol{X}$ and $\boldsymbol{Y}$ are arbitrary 3-vectors and $\boldsymbol{\sigma}$ is the Pauli spin matrix, we may write the free-electron
Schr\"odinger Hamiltonian as
\begin{equation}
\mathcal{H} = \frac{P^2}{2m_e} = \frac{(\bm{\sigma} \cdot \mathbf{P})^{2}}{2 m_e} 
\end{equation}
To include electromagnetic interactions we couple the vector potential $\mathbf{A}$ as prescribed by nonrelativistic
mechanics $(\mathbf{P} \rightarrow \bm{\Pi} = \mathbf{P}- e\mathbf{A}/c)$ and expand

\begin{equation}
\frac{(\bm{\sigma} \cdot \bm{\Pi})(\bm{\sigma} \cdot \bm{\Pi})}{2 m_e} 
\end{equation}
%
to obtain 
\begin{equation}\mathcal{H} \!= \!\frac{(\mathbf{P}- e\mathbf{A}/c)^2}{2m_e} + i\bm{\sigma}\cdot  \frac{(\mathbf{P}- e\mathbf{A}/c)\times  (\mathbf{P}- e\mathbf{A}/c)}{2m_e}. \label{eq:expanded}
\end{equation}
Then with the substitution $\mathbf{P} = -i\hbar\nabla$ and recognizing $\nabla \times \mathbf{A} = \mathbf{B}$
 as the magnetic field, we find
\begin{equation}
\bm{\Pi} \times \bm{\Pi}=\frac{i e \hbar}{c} \mathbf{B}.
\end{equation}
 The second term of \eq{eq:expanded} becomes
 \begin{equation}
 -\frac{e\hbar}{2m_e c}\medspace\bm{\sigma}\cdot\mathbf{B}, \label{eq:gfromS}
 \end{equation}
 so the gyromagnetic ratio $g=2$  emerges from the Schr\"{o}dinger equation, without specific mention of relativity. We did have to identify the electron as a spin-$\frac{1}{2}$ particle, whereas that emerges from Dirac's construction, and we all learned in school that spin-$\frac{1}{2}$ particles correspond to particular representations of the Lorentz group, which seems to point to relativistic roots~\cite{[{For an argument that spin emerges from (nonrelativistic) Galilean invariance, see }][]cmp/1103840281}.
 

\begin{acknowledgments}
I thank Aida El-Khadra, Andreas Kronfeld, and Mogens Dam for helpful correspondence,  Sam McDermott for calling my attention to Ref.~\cite{Beresford:2019gww}, and Florian Herren for pointing me  to Ref.~\cite{Fael:2013ij}.
Fermilab is operated by Fermi Research Alliance, LLC under Contract No. DE-AC02-07CH11359 with the U.S. Department of Energy, Office of Science, Office of High Energy Physics.
\end{acknowledgments}

\nocite{*}

\bibliography{Dmoment-v4}

\end{document}